\documentclass[twocolumn, superscriptaddress,pre,amsmath,amssymb, showpacs, footinbib]{revtex4}

\usepackage{graphicx}
\usepackage{dcolumn}
\usepackage{bm}

\begin{document}

\title{Dynamics of a colloid-stabilized cream}

\author{E.M.~Herzig}
\affiliation{SUPA School of Physics \& Astronomy, University of Edinburgh, Mayfield Road, Edinburgh, EH9 3JZ, United Kingdom}

\author{A.~Robert}
\affiliation{European Synchrotron Radiation Facility, Bo\^{i}te Postale 220, 38043 Grenoble, France}
\affiliation{LCLS, SLAC, 2575 Sand Hill Road, Menlo Park, CA 94025, USA}

\author{D.D.~van~'t~Zand}
\affiliation{SUPA School of Physics \& Astronomy, University of Edinburgh, Mayfield Road, Edinburgh, EH9 3JZ, United Kingdom}

\author{L.~Cipelletti}
\affiliation{Laboratoire des Collo\"{i}des, Verres et Nanomat\`{e}riaux, Universit\'{e} Montpellier 2 \& CNRS, 34095 Montpellier, France}

\author{P.N.~Pusey}
\affiliation{SUPA School of Physics \& Astronomy, University of Edinburgh, Mayfield Road, Edinburgh, EH9 3JZ, United Kingdom}

\author{P.S.~Clegg\footnote[1]{Corresponding author: paul.clegg@ed.ac.uk}}
\affiliation{SUPA School of Physics \& Astronomy, University of Edinburgh, Mayfield Road, Edinburgh, EH9 3JZ, United Kingdom}

\date{\today}

\begin{abstract}
We use x-ray photon correlation spectroscopy to investigate the dynamics of a high volume fraction emulsion creaming under gravity. The dodecane-in-water emulsion has interfaces stabilized solely by colloidal particles (silica). The samples were observed soon after mixing: as the emulsion becomes compact we discern two regimes of ageing with a cross-over between them. The young emulsion has faster dynamics associated with creaming in a crowded environment accompanied by local rearrangements. The dynamics slow down for the older emulsion although our studies show that motion is associated with large intermittent events. The relaxation rate, as seen from the intensity autocorrelation function, depends linearly on the wave vector at all times; however, the exponent associated with the line shape changes from 1.5 for young samples to less than 1 as the emulsion ages. The combination of ballisticlike dynamics, an exponent that drops below 1 and large intermittent fluctuations has not been reported before.
\end{abstract}


\maketitle

\section{Introduction}

Colloid-stabilized emulsions are metastable composites which can be remarkably robust. The interfacial colloids are trapped extremely strongly~\cite{Binks06} because they reduce the shared area between the dispersed and continuous phases. When an emulsion is initially prepared the droplets coalesce at a rate that decreases with increasing size corresponding to an increasing area fraction of interfacial colloids~\cite{Arditty03}. Ultimately a complete coverage of the interface by the particles provides a steric barrier against coalescence (although other mechanisms of stabilization are also possible~\cite{Vignati03}~\cite{Horozov06}~\cite{Leunissen07}). Unlike the interfaces of a conventional, surfactant stabilized, emulsions the densely packed colloids create a semi-solid layer~\cite{Clegg08}. This means that anisotropic stresses can be supported by these liquid-liquid interfaces~\cite{Subramaniam05}~\cite{Herzig07} and that a high volume fraction emulsion is a novel solid~\cite{LealCalderon08}. 

Here we report an experiment on colloid-stabilized emulsions as they become a compact cream layer; only preliminary studies currently exist~\cite{Yan97}~\cite{Binks99}. In our samples the typical final droplet sizes are on the scale of ten or more microns. The coalescence rates observed by Arditty and coworkers~\cite{Arditty03} strongly suggest that by the time our measurements begin the droplets have already reached full and stable coverage. Hence, this is a non-thermal system where gravity will drive robust droplets to form a high volume fraction cream: a process of solidification. For our preparation route there is a broad distribution of droplet sizes which implies that the cream will not tend to crystallize as the volume fraction increases: the likely end state of this system is a disordered random packing of droplets. The distribution also implies that there will be a spread of creaming rates that will necessarily lead to collisions between droplets from the earliest times. As described below, the emulsion begins at a volume fraction of $\Phi = 0.4$ and then becomes increasingly compact; from a very early stage during the experiment all the droplets are in contact. 

The creaming of conventional emulsions has been probed in great detail because it is of both fundamental and industrial importance~\cite{Robins00}. In conventional emulsions three generic types of behavior have been identified: I. If the emulsion is monodisperse with no inter-droplet attractions a close-packed cream layer can form while a clear solvent layer appears at the bottom. Between these two phases is a low volume fraction dispersion which is eradicated as the lower and upper interfaces move towards each other during creaming. II. If the emulsion is polydisperse the lower interface becomes diffuse rather than sharp while the volume fraction of the cream layer continuously increases. III. If the droplets have attractive interactions a gel can form which may slowly compactify or suddenly collapse. No low volume fraction dispersed phase is observed. In all three cases the compact cream layer can have some similarities to a liquid-liquid foam. The coarsening of foams has been studied using diffusive wave spectroscopy and it has been observed that ripening of bubbles leads to the build-up of stress followed by sudden avalanche rearrangements~\cite{Durian91}.
\begin{figure*}
\centerline{\includegraphics[scale=0.2]{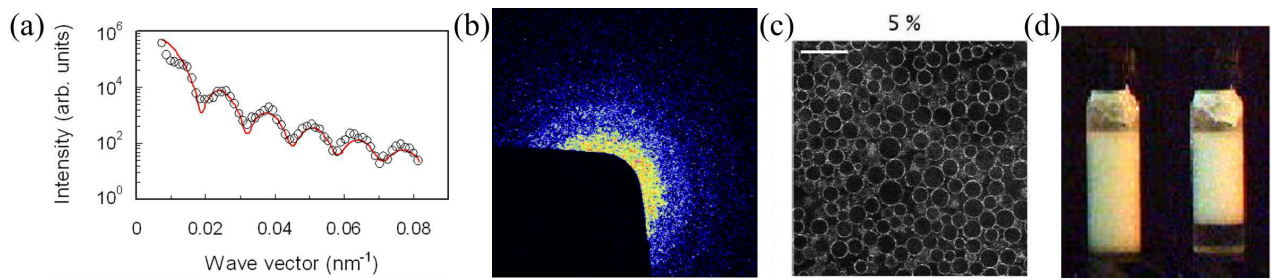}}
\caption{(Color online) (a) The form factor of the particles used in this experiment measured using small-angle x-ray scattering; (b) raw x-ray data showing a snap-shot of a speckle pattern; (c) confocal microscopy image of a particle-stabilized emulsion. Dodecane is the dispersed phase; the white layers are the silica colloids trapped at the interfaces. The scale bar is 50~$\mu$m; (d) photographs of an emulsion ($\Phi$ = 0.25 to ease visualization) illustrating creaming taking place.}
\label{photographs}
\end{figure*}

Our samples are opaque to visible light and contain constituents with high electron density contrast. In addition the dynamics are slow and we do not know prior to the experiment whether the samples will be ergodic. To explore these systems a powerful approach is to use X-ray Photon Correlation Spectroscopy (XPCS) using a charge-coupled detection device (CCD) as a two-dimensional detector. The samples are in the single-scattering regime for 8\,keV x-rays and, by recording the scattering at many pixels corresponding to the same wave vector, ensemble averaging can be carried out. Recently developed techniques can be used to reveal a detailed description of the fluctuations of the sample about its average behavior~\cite{Cipelletti05}~\cite{Duri05}.

Recent research on some soft-glassy materials using photon correlation spectroscopy both with visible light and with x-rays has shown a linear relationship between the probed length scale and the relaxation time in combination with faster than exponential relaxation. Furthermore, the dynamics are often characterized by intermittent events and show a delicate dependency on sample age~\cite{Cipelletti05}. This behavior is in stark contrast to Brownian dynamics but its origin is so far not fully understood. While a few soft glassy materials exhibit intermittent dynamics as a result of small temperature fluctuations~\cite{Mazoyer06} or for undiagnosed reasons~\cite{Guo07} most systems are responding to some driving force. An emulsion creaming under gravity is just such a system; again we find that the late-stage macroscopically quiescent system is remarkably active on a microscopic scale.

\section{Materials and methods}

The liquids forming the emulsion are water (Elga Purelab, 18.2\,M$\Omega$cm) and dodecane (Aldrich, $>$98\%) used without further purification. The anticipated interfacial tension of 72.8~mN/m at 20$^{\circ}$C~\cite{Janczuk89} leads to extremely strong trapping of the colloids used (more than 8~$\times$~10$^5$~k$_B$T per particle if the wetting angle of our particles lies between 60$^{\circ}$ to 120$^{\circ}$). Water-oil interfaces can become charged during emulsion preparation~\cite{Maeda04} and this can inhibit the trapping of colloids on interfaces. To suppress this effect we add KCl (Aldrich, 99+\%) sufficient to give a Debye screening length of 0.2~nm. The particles are silica prepared using the St\"{o}ber method~\cite{Stoeber68}. The particle radius, 243~$\pm$~7~nm, was determined by fitting the form factor for a sphere to the time-averaged small-angle x-ray data~\cite{Robert05} as shown in Fig.~\ref{photographs}(a). An ultrasound probe (Sonics \& Materials) is used to disperse the particles, 0.110 g (5\,vol\%) or 0.165 g (7.5\,vol\%), in 0.6~ml water. Then the dodecane (0.4 ml) and KCl (0.1~mg) are added before further mixing. A  syringe with long needle was used to fill a glass capillary (Hilgenberg). Some of the same sample was also examined using bright-field microscopy to confirm that an emulsion formed. 
\begin{figure*}
\includegraphics[scale=0.4]{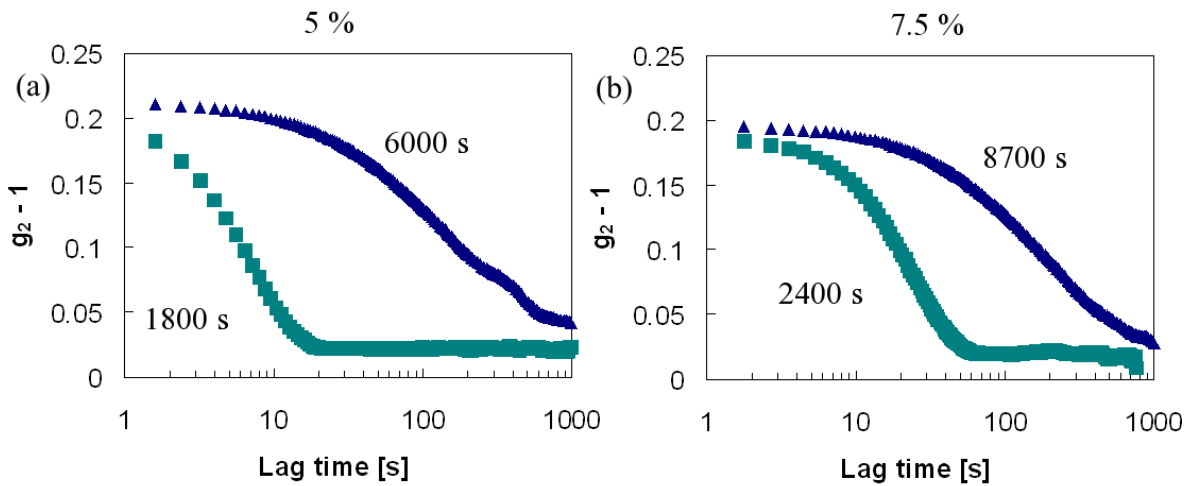}
\caption{(Color online) Intensity autocorrelation function calculated at $q = 0.033$~nm$^{-1}$ for two different ages as written on the graphs: early (squares) and late (triangles). Colloid volume fractions (a) 5\% (b) 7.5\%.}
\label{regimes}
\end{figure*}

The experiment was carried out at the Tro\"{i}ka beamline at the ESRF with an 7.979 keV x-ray beam ($\lambda = 0.155$~nm). We used a channel cut Si (111) monochromator ($\Delta \lambda / \lambda \approx 10^{-4}$) and a 10\,$\mu$m $\times$ 10\,$\mu$m aperture before the sample was used to create a partially coherent beam. The scattering from the colloid-stabilized emulsion was recorded using a CCD with 22.5\,$\mu$m pixel size located 2.3\,m after the sample (see Fig.~\ref{photographs}(b)) covering wave vectors 0.01~nm$^{-1}$ $< q <$ 0.08~nm$^{-1}$. Analyzing the intensity variations yields quantitative information about the dynamics occurring within the sample. This is found from the intensity autocorrelation function: $g_2(q, \tau)= \langle I(q,0) I(q, \tau)\rangle / \langle I(q) \rangle^2$ where $\langle \ldots \rangle$ denotes the average over a $\sim$20 minute time window and a ring of pixels sharing the same wave vector, $q$. The intermediate scattering function $f(q,\tau)$ is proportional to the square root of $g_2(q, \tau)-1$; we model this using a modified exponential function:
\begin{equation}
g_2(q, \tau) - 1 = b + c \exp{\left(-2\left(\frac{\tau}{\tau_c}\right)^\beta\right)}
\label{decay}
\end{equation}

\noindent with $b$ as the value to which the curve decays; $c$ is the apparent contrast which defines the difference in height between the initial and final level of the curve and $\tau_c$ is the relaxation time~\cite{f_note}. The function $g_2(q, \tau) - 1$ can be exponential ($\beta$=1), compressed exponential ($\beta >$1) or stretched exponential ($\beta <$1). Extracting parameters giving the best-fit between the model and the data for different $q$ allows us to establish a relationship between the length scale (i.e. inversely proportional to the wave vector), the characteristic relaxation time $\tau_c$ and the exponent $\beta$.  

We examine the evolution of the intensity autocorrelation function over time to establish whether ageing and intermittency are present. To do so the degree of correlation, $c_I(t, \tau)$ is calculated for a set of fixed lag times $\tau$ at a given wave vector $q$~\cite{Cipelletti05}.
\begin{equation}
c_I(t, \tau) = \frac{\langle I_p(t) I_p(t, \tau)\rangle _p}{\langle I_p(t)\rangle _p \langle I_p(t,\tau)\rangle _p} - 1,
\label{cI}
\end{equation}

\noindent where $I_p$ denotes the intensity at pixel $p$ and $\tau$ is the lag time between two frames while $\langle \ldots \rangle _p$ denotes the average over all pixels contained in a given annulus of the CCD image corresponding to a given $q$. It is possible to detect intermittent behavior in the $c_I(t, \tau)$ signal by examining the variation of $c_I(t, \tau)$ around its mean for different lag times~\cite{Duri05}. Homogeneous dynamics show a variance of $c_I(t, \tau)$ that varies with $c_I(t, \tau)$; however, for heterogeneous dynamics the variance will show a peak around the typical relaxation time i.e. the time when most of the intermittent events take place. This method will only resolve such a peak around $\tau_c$ if the fluctuations of the noise are smaller than the signal due to the particle dynamics.

\section{Results and Analysis}

Here we compare the dynamics and ageing of two oil-in-water emulsions containing $\Phi_v$ = 5\% and 7.5\% of silica colloids. As observed by confocal microscopy the typical droplet sizes are 16~$\pm$~4~$\mu$m and 12~$\pm$~3~$\mu$m diameter respectively, e.g. Fig.~\ref{photographs}(c). Measurements began as soon after emulsification as possible. The time-resolved speckle patterns were recorded for over 2 hours (with the samples returned later for subsequent measurements) and from these images the time-resolved correlation function $c_I(t, \tau)$ is calculated, Eq.~(\ref{cI}).  
\begin{figure*}
\includegraphics[scale=0.45]{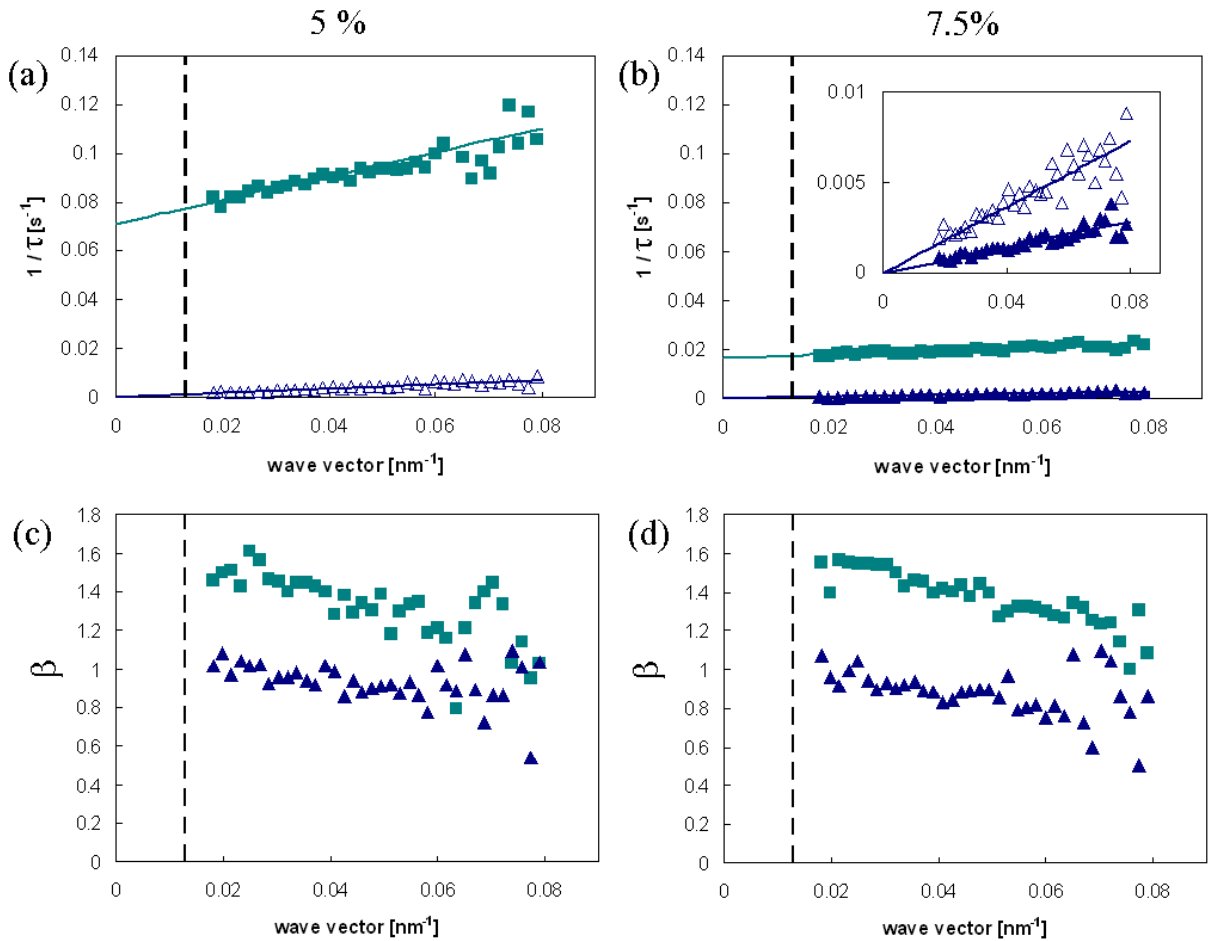}
\caption{(Color online) (a, b) Relaxation rate $1/\tau_c$ versus wave vector (c, d) exponent $\beta$ versus wave vector, see Eq.~(\ref{decay}): early (squares) and late (triangles). The dashed vertical line corresponds to the wave vector $q = 0.013$~nm$^{-1}$ which corresponds to $2\pi / d$, where $d$ is the diameter of the silica particles. (a, c) 5\%, (b, d) 7.5\%. Inset to (b): Relaxation rate at late times for both samples with an expanded y-scale.}
\label{parameters}
\end{figure*}

Figure~\ref{regimes} shows the change in $g_2(\tau) - 1$ with age for the two different samples at $q = 0.033$~nm$^{-1}$. A structure factor peak due to static particle-particle correlations would occur at $q \le 0.013$~nm$^{-1}$ which is at the lower limit of our wave vector range (marked by the dashed vertical line in Fig.~\ref{parameters}). Hence we are exploring the dynamics at or smaller than the particle scale.
 
The emulsion behavior can be divided into two regimes with a cross-over between them. Initially the relaxation time is short, then there is a transition where the system starts slowing down and in the final regime it reaches a significantly longer relaxation time. For the 5\% sample (larger droplets) the initial decay is significantly faster than that of the 7.5\% sample (smaller droplets). The onset of the slowing down occurs in both cases after about 1 hour 20 minutes but is drawn out over almost an hour for the small droplets while the change from fast to slow relaxation happens over 10 minutes for the 5\% sample. The final slow regime in the 5\% sample reaches an initial plateau; however, significant fluctuations in the relaxation time are observed over the remaining time for which measurements were made. The slower ageing evolution of the 7.5\% sample leads to only a short segment of the final stage being observed; however, this still seems to plateau at a similar relaxation time as the 5\% sample.

Shown in Fig.~\ref{photographs}(d) is a more dilute emulsion ($\Phi$ = 0.25) before and after creaming for one hour; the cream layer occupies well over half of the sample volume. For the x-ray studies, $\Phi$ = 0.4, giving a cream layer that occupies most of the sample volume: the x-ray beam passes through the cream layer at all times and for all samples. Repeat measurements were made after the sample capillary had been removed from the instrument and later remounted (data not shown). In some cases the disturbance to a late stage sample led to a return to early stage behavior even though the sample itself was then hours or days old. Since the coalescence of colloid-stabilized droplets is completed within minutes~\cite{Arditty03} we do not associate the early stage regime with coalescence. Instead it appears likely that early stage behavior can be a consequence of disrupting the droplets as an inevitable part of mounting the sample at the beamline.
\begin{figure*}
\includegraphics[scale=0.5]{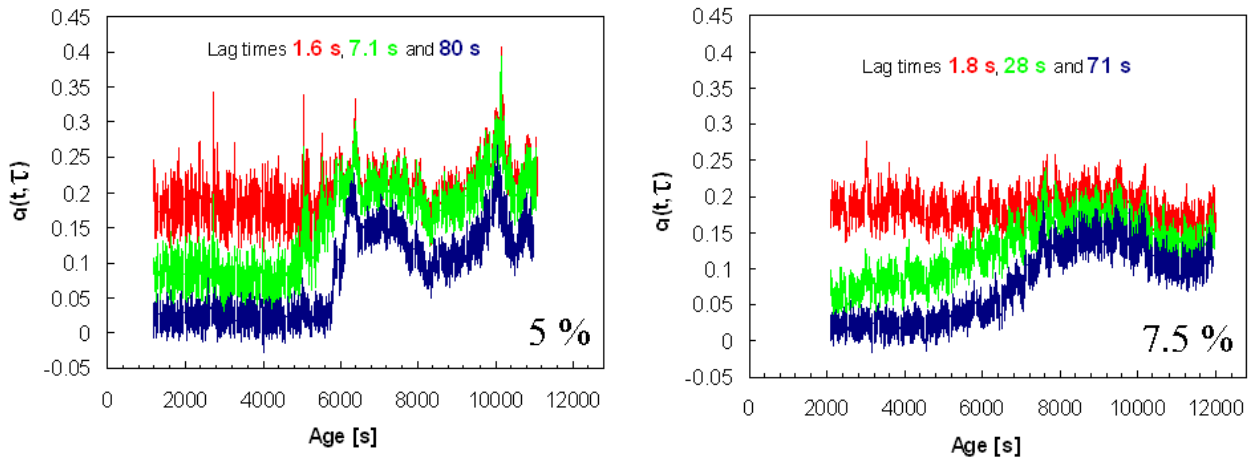}
\caption{(Color online) The signal $c_I(t, \tau)$ at $q = 0.033 \pm 0.003$~nm$^{-1}$ for different $\tau$ values plotted against sample age shows that large fluctuations are present. (a) 5\%, (b) 7.5\%, lag times given for curves from top to bottom.}
\label{raw_cI}
\end{figure*}

Qualitatively, the intensity autocorrelation function (Fig.~\ref{regimes} for $q = 0.033$~nm$^{-1}$) shows a significant increase in the relaxation time with age for both volume fractions. At the same time it is apparent that the shape of the decorrelation changes while the sample ages. To quantify these effects the best fit of Eq.~(\ref{decay}) to all the obtained curves was found. The optimal fit parameters for all wave vectors are shown in Fig.~\ref{parameters}. For diffusive dynamics we expect the relaxation rate $1/\tau_c \sim q^2$ and $\beta = 1$; for ballistic motion we expect $1/\tau_c \sim q$ with a value of $\beta$ which reflects the velocity distribution~\cite{Cipelletti03} ($\beta = 2$ corresponds to a Gaussian distribution). Evidently, $1/\tau_c$ varies linearly with $q$ for both samples and at all times showing that the dynamics leading to the relaxation are not diffusive; however, there is a significant change between the behavior at early and late times. For early times there is a sizeable off-set such that $1/\tau_c$ does not decrease to zero for small $q$. This is observed for both volume fractions and is also evident in other data sets not presented here. In addition to this off-set the slope of the $1/\tau_c$-curve decreases with time as the droplets are compactified (the slope is smaller for the smaller droplets, Fig.~\ref{parameters}(b) inset).

The background level, $b$, to which the intensity autocorrelation function decays, eq.~(\ref{decay}), shows slight wave-vector dependence. We observe non-zero values for $g_2(q, \tau) - 1$ at long times for $q$ values close to the maxima of the form factor of the particles. Non-zero $g_2(q, \tau) - 1$ values at long times are typically associated with non-ergodic sample dynamics; however, associating error bars with these long time measurements is challenging. In the early stage the existence of a component of $1/\tau_c$ which is independent of wave-vector implies that the samples have dynamics that lead to complete decorrelation at a steady rate. The long-time behavior is hence not due to non-ergodicity but instead appears to be a result of the small size of the signal. For late stage creaming behavior, where the dynamics are slow, there are insufficient data to firmly determine the long-time behavior of the sample. 

Although the time scales change for emulsions stabilized using two different volume fractions of colloids the overall aging behavior does not. At early times the exponent $\beta$ has values around 1.5 for smaller wave vectors probed (see Fig.~\ref{parameters}(c) and (d)) and slowly decreases towards 1.0 for the larger wave vectors~\cite{Duri06}. At later times $\beta$ is smaller starting at 1.0 for lower wave vectors and decreasing further for the largest wave vectors probed. The values imply that relaxation is faster than exponential (compressed) at least at early times. Such behavior, which has also been found by numerous other researchers in a wide variety of samples, cannot be created from the sum of exponential processes and its interpretation is the subject of great current activity~\cite{Cipelletti05}~\cite{Duri05}~\cite{Mazoyer06}~\cite{Guo07}~\cite{Cipelletti03}~\cite{Duri06}~\cite{Trappe07}~\cite{Caronna08}~\cite{Bouchaud01}~\cite{Tanaka07}. At late times the intensity autocorrelation function is a slightly stretched exponential: again this is not typically seen in concert with ballisticlike relaxation rate behavior.

To determine whether the average dynamics capture what is essential about this system we need to establish if the rearrangements under gravity are continuous over time, representing a slow pushing of droplets together, or instead the upward drive induces a multitude of separate, small rearrangements. To carry out this analysis an optimized $q$-ring (large number of pixels, high intensity) was chosen and time sequences are selected (as described below) to calculate the variance of $c_I(t, \tau)$ over a specific time range for the measured $\tau$-range. Figure~\ref{raw_cI} shows the fluctuations of $c_I(t, \tau)$ for different values of $\tau$ over the measurement time of the sample. The two regimes of sample dynamics are clear in these raw data. By calculating $c_I(t, 0)$ which by definition is a static value~\cite{Duri05} and examining its fluctuations it is evident that noise forms a substantial fraction of some time sequences. It is not possible to directly correct for this noise using the number of pixels available. Instead, segments of $\sim$20 minute duration, in which the noise seen in $c_I(t, 0)$ is less significant, are chosen for determining the variance of the $c_I(t, \tau)$-signal. A large variance will indicate that an important role is being played by more extreme events.
\begin{figure*}
\includegraphics[scale=0.45]{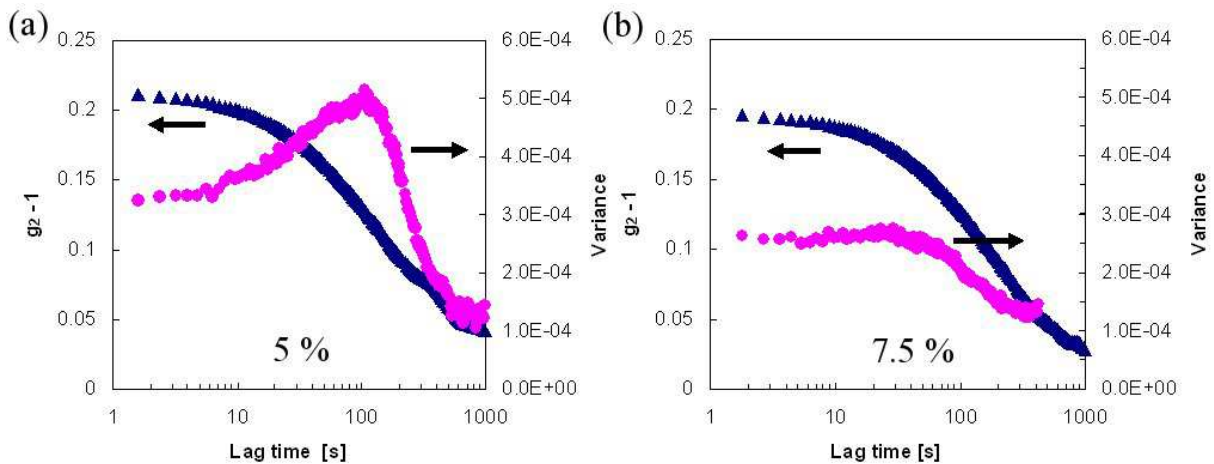}
\caption{(Color online) A peak in the variance in the region of the relaxation time is an indication for intermittent behavior. Triangles: $c_I(t, \tau)$ for optimized $q$-range and time segment (late stage) that does not show strong photon fluctuations, circles: variance of $c_I(t, \tau)$ for different $\tau$. (a) 5\% and (b) 7.5\%.}
\label{variance}
\end{figure*}

Figure~\ref{variance}(a, b) shows the time averaged correlation functions $g_2(q, \tau)-1$ for 5\% and 7.5\% with the variance of the underlying $c_I(t, \tau)$ signal over time $t$ at each $\tau$ overlaid. Especially for the late 5\% sample a clear peak around the relaxation time is visible suggesting that intermittent rearrangements occur at late stages. Less of the late stage was accessible for the 7.5\% sample. For the early times (data not shown) it does not appear that the variance shows a peak but due to the lack of data for small enough $\tau$ this cannot be said for certain. There is no indication that these intermittent events are associated with a collapse of the emulsion; it is most likely that our emulsions are type II.

\section{Discussion}

In these experiments the x-rays are scattered from colloidal silica, irreversibly trapped at the surface of oil droplets. The experimentally accessible wave vector range corresponds to the colloidal scale rather than the droplet scale. Hence our samples are organized into large units with correlated dynamics on the scale we are observing. The droplets are driven to rise and pack more densely by gravity and the colloids are carried along as this process takes place. 

Because a linear variation of relaxation rate with wave vector is a signature of ballisticlike dynamics we need to begin by considering how fast we would expect creaming to occur. Yan and Masliyah have applied the concepts of hindered creaming to particle-stabilized emulsions~\cite{Yan97}. The analysis is based on the model of hindered settling due to Richardson and Zaki: the creaming velocity is expected to be $v_c = v_0 (1 - \Phi)^n$ where $v_0$ is the creaming velocity of the unhindered subunits and the exponent $n$ depends on the interactions between the subunits (with a value around 5 being typical)~\cite{Richardson54}~\cite{Davies76}. For our droplets at volume fraction $\Phi = 0.4$, using $n=5$, this analysis gives a hindered creaming velocity of $v_c \approx 300$~nm/s. 

The early stage creaming under gravity is characterized by a relaxation that is independent of wave vector as indicated by the intersection of the $1/\tau_c$-line with the y-axis (see Fig.~\ref{parameters}(a) and (b)). From the intercept we extract the rate of `whole sample changing' events. The 5\% and 7.5\% samples fully decorrelate at quite a high rate (every 14 and 60 seconds respectively at this age). The time it takes for the creaming sample to pass through the 10~$\mu$m beam (about 30~sec) according to the Richardson and Zaki model is roughly right to account for these decorrelations. The difference in the rate of `whole sample changing' events between the two emulsion compositions is larger than predicted by the model of hindered settling. We note that the hindered creaming velocity depends on the volume fraction of the emulsion to a high power (here taken to be 5). Small changes in the composition of the emulsion transferred to the x-ray capillary will be reflected in large changes in the creaming velocity. It is likely that this accounts for the differences observed.

Upward sample motion accounts for the wave vector independent aspect of the relaxation rate; the linear $q$-dependent relaxation process \textit{rides on the back} of the steady creaming and is presumably associated with local rearrangements and possibly droplet rotation. This behavior can be contrasted with the complex relaxation of gels observed by Trappe and coworkers~\cite{Trappe07}. Here $q$-dependent and $q$-independent relaxations were observed simultaneously i.e. the intensity autocorrelation function exhibited two separate shoulders: one with a $\tau_c$ that varied with $q$ and the other with a different $\tau_c$ which did not vary with $q$. Hence separate ballisticlike and large-scale rearrangements were taking place within these gel samples. In our study the creaming and ballisticlike rearrangements are combined in a single relaxation process.  

The transition to the late stage is characterized by a marked change in the sample dynamics (possibly corresponding to a transition from the dispersed droplet region to a dense cream~\cite{Robins00}). At the simplest level the sample relaxation slows down; however there are four other features involved. First, the $q$-independent aspect of the dynamics comes to an end. This is evident in the decay rate versus wave vector graph meeting the y-axis at the origin (see inset to Fig.~\ref{parameters}(b)) and implies that the emulsion is becoming compact. Second, for these droplet sizes and volume fractions we find that the variation of the decay rate with wave vector is independent of the orientation of the wave vector; such isotropy is common to high volume-fraction granular systems settling under gravity~\cite{Nicolai95}. (In studies of sedimentation velocities using direct imaging of particles at lower volume fractions, $\Phi \le 0.2$, show sustained anisotropy in the velocity and its variance~\cite{Segre97}~\cite{Nicolai95a}.) Third, the exponent $\beta$ which describes the dynamic structure factor falls to one or below. Fourth, the dynamics appear to be characterized by a large variance which peaks at the characteristic time of the system relaxation. Hence the system combines ballisticlike dynamics with intermittency (which has been observed previously for various systems) with a stretched exponential intermediate scattering function (which is unprecedented). The growth in the variance of fluctuations is likely to be a consequence of the increasingly correlated behavior of the dense cream.

Often, soft glassy systems have dynamics that slow down with age. Studies have been carried out for dense colloids, shaving foam, surfactant `onions' and many others. Cipelletti et al.~\cite{Cipelletti03} and others~\cite{Tanaka07} suggest that the characteristic dynamics are due to the build-up of internal stresses. The slowing down is then due to either a decrease in the rate of change of the stress source strength or a reduction in the number of active stress sources as the sample ages. In our case it is possible that the early regime is initiated either by formation or by the loading of the sample; this is strongly supported by the fact that it can (sometimes) be re-induced by removing the sample and replacing it again later. Hence, the two regimes we observe in the dynamics of particle-stabilized emulsions share much in common with the two regimes observed by Guo and coworkers in the dynamics of nanoemulsions~\cite{Guo07}. Nanoemulsions are stabilized using conventional surfactants and have subunits of much smaller size. It was observed using XPCS that during the period immediately following the loading of the sample the dynamics were fast; however, this behavior was observed to give way after 10-20 minutes to a regime characterized by slower dynamics. The initial fast dynamics were interpreted as being due to an initial population of stress dipoles created during the loading of the sample; the drive behind the dynamics in the slower regime was not unambiguously identified. In our case the initial phase is likely to have dynamics that are driven both by the population of stress dipoles created during loading and also by the creaming under gravity. The drive due to gravity will not disappear in the second regime that we observe and this is likely to give rise to the late-stage dynamics.

The linear $q$-dependence of $1/\tau_c$ such as that observed here in both regimes (Fig.~\ref{parameters}) has been attributed to stress dipoles within samples. It implies that the direction of rearrangements within the sample is maintained across several rearrangements. Considering rearrangement events as randomly distributed in time Duri and Cipelletti~\cite{Duri06}~\cite{Bouchaud01} have shown that if they are dipolar stresses then this necessarily leads to a $\beta$-exponent of 1.5 for low wave vector that approaches 1.0 for large wave vector. Our observations yield the same trend in $\beta$ in the early stage; although, all of our measurements are carried out at large wave vector. For the late stage where intermittent events appear to dominate the stress source is still gravity pushing the particles upwards into each other. This no longer gives $\beta$= 1.5. In all experimental observations to date the intermediate scattering function accompanying the ballisticlike dynamics could be described using a compressed exponential where the compression exponent was a little larger than 1. Recent experimental studies have shown that the dynamics are characterized by intermittent events~\cite{Duri06}~\cite{Trappe07}. The results presented here differ in one key respect: the late-stage dynamic structure factor is stretched rather than compressed. Hence we have observed experimentally that ballisticlike dynamics together with large intermittent rearrangements do not necessarily have a faster than exponential relaxation. Stretched exponential behavior does not emerge from the model of Bouchaud and Pitard~\cite{Bouchaud01}. An intermediate scattering function with stretched exponential form is often observed for colloidal gel samples especially for measurements at low wave vector~\cite{Zaccarelli07}. Here the dynamics are  associated with structural heterogeneity on the scale of clusters of colloids. In our experiments the dynamics have been characterized on the particle scale and so the same explanation is unlikely to apply. The compressed exponential relaxation of Bouchaud and Pitard~\cite{Bouchaud01} was associated with a long-range elastic deformation field; the model predicts that the exponent will fall at larger wave vectors (and this has been observed). However an exponent of less than 1 does not sit easily with this interpretation. 

\section{Summary}

We have carried out an XPCS study of colloid-stabilized emulsions as they cream and form a novel soft solid. The data show two distinct regimes and a cross-over between them: the early time regime is characterized by a creaming of the droplets combined with local rearrangements. As the samples age a peak in the variance of the correlation function is observed (most clearly for $\Phi_v$ =5\%) with the peak located at the characteristic relaxation time. This strongly suggests that the droplets undergo intermittent rearrangements during compactification. The increasing dominance of large fluctuations suggests that large regions of the sample are becoming strongly correlated. The form of the intermediate scattering function is unusual for systems exhibiting ballisticlike dynamics with an intermittent character. Large, colloid-stabilized, droplets form intriguing systems for further study: the ability to select the size of the irreversibly trapped particles provides a means to control the roughness of granular scale objects.

\section{Acknowledgments}
We are grateful to W.~Poon for helpful comments and to A.~Schofield for preparing the particles. Funding was provided by the EPSRC (EP/D076986/1) and by the EU NoE SoftComp. LC is a member of the Institut Universitaire Francais, whose support is gratefully acknowledged. The measurements were carried out at the Tro\"{i}ka beamline ID10A of the European Synchrotron Radiation Facility, Grenoble (France).


\begin{thebibliography}{100}
\bibitem{Binks06}
B.P.~Binks and T.S.~Horozov \textit{Colloidal particles at liquid interfaces} (Cambridge University Press, Cambridge, 2006) pp.~1--74.
  
\bibitem{Arditty03}  
S.~Arditty, C.P.~Whitby, B.P.~Binks, V.~Schmitt and F.~Leal-Calderon, Eur. Phys. J. E \textbf{11}, 273 (2003).

\bibitem{Vignati03}
E.~Vignati, R.~Piazza and T.P.~Lockhart, Langmuir \textbf{19}, 6650 (2003).

\bibitem{Horozov06}
T.S.~Horozov and B.P~Binks, Angew. Chem. Int. Ed. \textbf{45}, 773 (2006).

\bibitem{Leunissen07}
M.E.~Leunissen, A.~van Blaaderen, A.D.~Hollingsworth, M.T.~Sullivan and P.M.~Chaikin, Proc. Natl. Acad. Sci. USA \textbf{104}, 2585 (2007).

\bibitem{Clegg08}
P.S.~Clegg, J. Phys.: Condens. Matter \textbf{11}, 113101 (2008).

\bibitem{Subramaniam05}
A.B.~Subramaniam, M.~Abkarian, L.~Mahadevan and H.A.~Stone, Nature \textbf{438}, 930 (2005).

\bibitem{Herzig07}
E.M.~Herzig, K.A.~White, A.B.~Schofield, W.C.K.~Poon and P.S.~Clegg, Nature Mater. \textbf{6}, 966 (2007).

\bibitem{LealCalderon08}
F.~Leal-Calderon and V.~Schmitt, Curr. Opin. Colloid Interf. Sci. \textbf{13}, 217 (2008).

\bibitem{Yan97}
N.~Yan and J.H.~Masliyah, Ind. Eng. Chem. Res. \textbf{36}, 1122 (1997).
  
\bibitem{Binks99}
B.P~Binks and S.O~Lumsdon, Phys. Chem. Chem. Phys. \textbf{1}, 3007 (1999)
  
\bibitem{Robins00}
M.M.~Robins, Curr. Opin. Colloid Interf. Sci. \textbf{5}, 265 (2000).
  
\bibitem{Durian91}
D.J.~Durian, D.A.~Weitz and D.J.~Pine, Science \textbf{252}, 686 (1991).

\bibitem{Cipelletti05}
L.~Cipelletti and L.~Ramos, J. Phys.: Condens. Matter \textbf{17}, R253 (2005).

\bibitem{Duri05}
A.~Duri, H.~Bissig, V.~Trappe and L.~Cipelletti, Phys. Rev. E \textbf{72}, 051401 (2005).

\bibitem{f_note}
This yields the same $\beta$ and $\tau_c$ values as modeling the intermediate scattering function, $f(q,\tau)$, using Eq.~(\ref{decay}) without the 2 in the argument of the exponential.
  
\bibitem{Mazoyer06}
S.~Mazoyer, L.~Cipelletti and L.~Ramos, Phys. Rev. Lett. \textbf{97}, 238301 (2006).
  
\bibitem{Guo07}
H.~Guo, J.N.~Wilking, D.~Liang, T.G.~Mason, J.L.~Harden and R.L.~Leheny, Phys. Rev. E \textbf{75}, 041401 (2007).
  
\bibitem{Janczuk89}
B.~Ja\'{n}csuk, T.~Bailopiotrowicz and W.~W\'{o}jcik, Colloids and Surfaces \textbf{36}, 391 (1989).

\bibitem{Maeda04}
N.~Maeda, K.J.~Rosenberg, J.N.~Israelachvili and R.M.~Pashley, Langmuir \textbf{20}, 3129 (2004).

\bibitem{Stoeber68}
W.~St\"{o}ber, A.~Fink and E.~Bohn, J. Colloid Int. Sci. \textbf{26}, 62 (1968).
  
\bibitem{Robert05}
A.~Robert, J.~Wagner, T.~Autenrieth, H\"{a}rtl and G.~Gr\"{u}ber, J. Chem. Phys. \textbf{122}, 084701 (2005).

\bibitem{Cipelletti03}
L.~Cipelletti, L.~Ramos, S.~Manley, E.~Pitard, E.E~Pashkovski, D.A.~Weitz and M.~Johansson, Faraday Discuss. \textbf{123}, 237 (2003).

\bibitem{Duri06}
A.~Duri and L.~Cipelletti, Europhys. Lett. \textbf{76}, 972 (2006).
  
\bibitem{Trappe07}
V.~Trappe, E.~Pitard, L.~Ramos, A.~Robert, H.~Bissig and L.~Cipelletti, Phys. Rev. E \textbf{76}, 051404 (2007).

\bibitem{Caronna08}
C.~Caronna, Y.~Chushkin, A.~Madsen and A.~Cupane, Phys. Rev. Lett. \textbf{100}, 055702 (2008).

\bibitem{Bouchaud01}
J.-P.~Bouchaud and E.~Pitard, Eur. Phys. J. E \textbf{6}, 231 (2001).

\bibitem{Tanaka07}
H.~Tanaka and T.~Araki, Europhys. Lett. \textbf{79}, 58003 (2007).  
  
\bibitem{Richardson54}
J.F.~Richardson and W.N.~Zaki, Trans. Inst. Chem. Eng. \textbf{32}, 35 (1954).
  
\bibitem{Davies76}
L.~Davies, D.~Dollimore and J.H.~Sharp, Powder Technol. \textbf{13}, 123 (1976).

\bibitem{Nicolai95}
H.~Nicolai, B.~Herzhaft, E.J.~Hinch, L.~Oger and E.~Guazzelli, Phys. Fluids \textbf{7}, 12 (1995).
  
\bibitem{Segre97}
P.N.~Segr\'{e}, E.~Herbolzheimer and P.M.~Chaikin, Phys. Rev. Lett. \textbf{79}, 2574 (1997).
  
\bibitem{Nicolai95a}
H.~Nicolai and E.~Guazzelli, Phys. Fluids \textbf{7}, 3 (1995).

\bibitem{Zaccarelli07}
E.~Zaccarelli, J. Phys.: Condens. Matter \textbf{19}, 323101 (2007).
\end{thebibliography}
\end{document}